\documentclass[aps,preprint,amsmath,amssymb,amsfonts,showpacs]{revtex4}
\usepackage{epsfig}
\usepackage{graphicx}
\usepackage{subfigure}
\usepackage{dcolumn}
\usepackage{bm}

\newcommand{\del}{\partial}

\newcommand{\ddsimple}[2]{\frac{\del #1}{\del #2}} 

\renewcommand{\Im}{\operatorname{Im}}

\newcommand{\boldk}{\mathbf{k}}

\begin{document}

\title{On-shell actions with lightlike boundary data}

\author{Yasha Neiman}
\email{yashula@gmail.com}
\affiliation{Institute for Gravitation \& the Cosmos and Physics Department, Penn State, University Park, PA 16802, USA}

\date{\today}

\begin{abstract}
We argue that finite-region observables in quantum gravity are best approached in terms of boundary data on null hypersurfaces. This has far-reaching effects on the basic notions of classical and quantum mechanics, such as Hamiltonians and canonical conjugates. Such radical properties are not unexpected in finite-region quantum gravity. We are thus motivated to reformulate field theory in terms of null boundary data. As a starting point, we consider the on-shell action functional for classical field theory in finite null-bounded regions. Closed-form results are obtained for free scalars and for Maxwell fields. The action of classical gravity is also discussed, to the extent possible without solving the field equations. These action functionals exhibit non-locality and, in special cases, a ``holographic'' reduction of the degrees of freedom. Also, they cannot be used to define global charges. Whereas for ordinary field theory these are just artifacts of a restrictive formalism, in quantum gravity they are expected to be genuine features. This further supports a connection between quantum gravity and null-boundary observables. In our treatment of the GR action, we identify a universal imaginary term that reproduces the Bekenstein entropy formula.
\end{abstract}


\maketitle
\tableofcontents
\newpage

\section{Introduction} \label{sec:intro}

Understanding quantum gravity (QG) in finite spacetime regions is the greatest conceptual problem facing physics today. The observed positive value of the cosmological constant implies that it cannot be sidestepped by resorting to experiments at infinity, since infinity is not accessible. In this paper, we advocate an approach to the issue that involves restricting to \emph{null-bounded} finite regions, i.e. to boundary data on lightlike hypersurfaces. 

The paper is composed of two parts. In the first part (section \ref{sec:why_null}), we present an abstract argument in favor of null boundary data. We list some conceptual difficulties that arise with respect to finite-region QG, and label them as either ``hard'' or ``secondary''. The ``hard'' issues have to do directly with the finite information bound \cite{Bekenstein:1972tm,Bekenstein:1973ur,Bekenstein:1980jp} implied by black hole thermodynamics. This renders obsolete the notion of infinitely precise observables \cite{Witten:2001kn}, a situation that no one knows how to deal with. In particular, it's likely that quantum mechanics should be modified, since at present it deals with infinitely precise real numbers, such as probabilities and couplings. 

In contrast, the ``secondary'' difficulties do not directly touch on the issue of finite precision, though they are sometimes conflated with it. An example is the issue of invariance under transverse diffeomorphisms. Our goal in section \ref{sec:why_null} is to show that the ``secondary'' problems can be removed by restricting the discussion to null boundary data. In this way, we propose to disentangle them from the ``hard'' problem, making the latter easier to approach. The bulk of our reasoning doesn't favor a particular type of QG theory, apart from the use of null boundaries. Nevertheless, in section \ref{sec:why_null:discrete} we make several arguments in favor of discrete geometries, in particular in the context of black hole creation and evaporation.

In the second part of the paper (sections \ref{sec:class}-\ref{sec:gravity}), we take baby steps in the proposed direction. Quantum gravity aside, the notion of restricting field theory to null boundary data has far-reaching consequences. We set out to explore these on the safe ground of classical field theory. In classical physics, the dynamics can be encoded in the on-shell action functional. In quantum theory, the (effective) on-shell action is upgraded into transition amplitudes, from which, again, all observables can be derived. The off-shell action, on the other hand, is upgraded into contributions to the path integral. These can be rearranged, and are therefore less physical than the overall amplitudes. Our strategy, then, is to treat the on-shell action as the basic ``observable'' in classical field theory, and to ask what features of the system can be derived from it. We will find that certain basic properties of field theory are no longer manifest in the null setup. The reason is that configuration variables at different points of a closed boundary are no longer independent. The resulting breakdown of standard features of field theory is potentially exciting, for reasons we will now discuss. For a similar discussion, see section 8 of \cite{ArkaniHamed:2008gz}.

With each conceptual step in physics, old certainties are exposed as either approximate or observer-dependent. In the process, the number of genuine, invariant features of the world is reduced. Knowing which concepts are lost at each step can be a valuable guide towards the new physics. In particular, it pays to find a description of the \emph{old} physics in which the doomed concepts are no longer obvious. A good example is the minimal-action formulation of classical mechanics. It makes determinism non-obvious, and in doing so paves the way towards quantum mechanics. Another example is the pre-symplectic formulation of Hamiltonian mechanics, in which time loses its special role. 

At present, we are in search of a conceptual framework for quantum gravity. There is a list of features of known physics that are expected to fade away in a QG world. First and foremost, space is expected to lose its infinite information-carrying capacity: black hole thermodynamics implies that the number of bits in a finite region is also finite, and proportional to the surface area. 

One also expects a loss of \emph{locality}. This appears to be a necessary result of the information bound \cite{ArkaniHamed:2007ky}, and is strongly implied by a holographic picture of the degrees of freedom \cite{'tHooft:1999bw,Susskind:1994vu}. At the very least, the accuracy with which one can measure a would-be pointlike observable now depends on the overall surface area of the experimentally available region. Note that this notion of non-locality is stronger than a simple Planck-scale resolution limit: it implies influences over an arbitrarily large distance. Also, it is of course stronger than the classical notion of ``no pointlike observables'' in gravity. The latter is due to the lack of a background geometry in which pointlike objects can be located. That problem can be resolved by restricting to observables on the same hypersurface as the boundary data (which is not necessarily at the boundary of spacetime). These observables are then local, in the sense that GR is a local field theory; we will elaborate on this in section \ref{sec:class:scalar_locality}.

Another notion that must apparently be abandoned in QG is that of conserved global charges \cite{Banks:2010zn,Kallosh:1995hi}. One way of stating the argument is that a black hole of fixed size can have an arbitrary value of a would-be global charge. This constitutes an unbounded amount of information, contradicting the holographic bound. 

It is therefore desirable to find a formulation of known physics in which bulk degrees of freedom, locality and global charges - the features that are expected to disappear in QG - are no longer manifest. This is precisely what happens when we examine classical field theory in terms of on-shell actions in null-bounded regions (though our exploration is in no way a complete \emph{reformulation} of classical field theory). Generically, we find that manifest locality is lost, and that global charges can no longer be defined. Also, in special cases such as free scalars and pure gravity, we find an apparent ``holographic'' reduction in the degrees of freedom. On a broader level, these properties can all be traced to a failure of the usual machinery of classical mechanics. This parallels the failure of quantum mechanics that is expected in finite-region QG. We stress that all of the above takes place for classical field theories, where it is merely an artifact of a restrictive formalism. Nonetheless, together with the general arguments from section \ref{sec:why_null}, these parallels build up a case for null-boundary observables as the correct approach to QG in finite regions. 

As a curious spin-off, our investigation of the null-bounded GR action in section \ref{sec:gravity} reveals an imaginary contribution (eq. \eqref{eq:Im_S}) to the Gibbons-Hawking boundary term. This contribution reproduces the Bekenstein-Hawking formula for black hole entropy \cite{Hawking:1974rv}, both in pure GR and in the presence of matter. Imaginary actions in GR are not a new notion, dating back to the analysis of black holes through Wick rotation \cite{Gibbons:1976ue}. Our derivation reveals this feature in Lorentzian language, with no restriction to stationary metrics. An imaginary action term is in fact present for generic boundaries in Lorentzian spacetime. However, its locality structure makes it irrelevant for action variations, in all but the null case. The relevant notions of locality are discussed in section \ref{sec:class:scalar_locality}.   

Section \ref{sec:discuss} contains a discussion and outlook on the classical results, along with some comments on the content of the quantum theory.

\section{Why finite null boundaries for quantum gravity} \label{sec:why_null}

In this section, we motivate the use of finite null-bounded regions in quantum gravity. This motivation was the starting point of the present work. 

\subsection{The necessity and challenge of finite regions} \label{sec:why_null:finite}

Through supergravity, string theory and AdS/CFT, various aspects of quantum gravity have become accessible. In particular, AdS/CFT \cite{Maldacena:1997re,Aharony:1999ti} provides a handle on non-perturbative dynamics. However, this understanding remains limited to observables at infinity. There is not even a clear definition of observables for a QG experiment in a local region of spacetime. Indeed, the entropy-bound arguments mentioned in the Introduction suggest that a paradigm shift is necessary before such experiments can be properly discussed. Due to the finite amount of information in a region with given surface area, it seems that the very notion of an observable (or a quantum-mechanical probability) as an infinitely precise real number must be discarded.

One may think that this problem is not very serious. Indeed, physics always deals with idealizations. So perhaps in quantum gravity, an ideal, well-defined experiment can only be performed at infinity. Then realistic, finitely extended experiments are understood as approximations, as they already are in many other ways. However, this reasoning collapses in the face of a positive cosmological constant. Just as $\hbar$ provides a fundamental, finite limit on the accuracy of classical physics, so does $\Lambda^{-1}$ provide a finite limit on the size of causally available spatial regions. The entropy available to any observer is then limited by the de-Sitter horizon area \cite{Banks:2000fe,Bousso:2000nf}. Arbitrarily precise observables at infinity no longer provide a valid idealization of real experiments, and the hard issues concerning finite regions must be addressed.

\subsection{``Secondary'' problems with finite regions} \label{sec:why_null:secondary}

Discrete QG models, such as the spinfoam formulation \cite{FK,EPRL,BC} of loop quantum gravity (LQG) \cite{Rovelli:2008zza}, are at first sight well-suited to address dynamics in finite regions. However, the physical meaning of e.g. spinfoam amplitudes is far from trivial. The relevant difficulties can be formulated in a model-independent way. These difficulties are sometimes seen as ruling out any ``simple-minded'' attempt to discuss finite-region QG. 

The first and foremost problem arises from the freedom of transverse diffeomorphisms, i.e. displacements of the boundary in perpendicular to itself. As a result of this redundancy, different boundary data may correspond to different slices of the same history, i.e. to the same state. In canonical GR, this redundant degree of freedom is generated by the Hamiltonian constraint, which also encodes the dynamics. The ``physical'' state space is obtained from the ``kinematical'' space of codimension-1 boundary data by imposing the Hamiltonian constraint, i.e. by solving the theory. While this doesn't invalidate spinfoam-like approaches, it means that the theory's formulation is only distantly related to its physical content, i.e. to the structure of the true state space and observables on it. This is the strong form of the statement that ``local observables are not diff-invariant''. The weak form, concerning \emph{pointlike} observables, arises already in classical GR, and is addressed by restricting to observables on the boundary-data hypersurface.

If we consider boundary data on finite \emph{spacelike}-bounded regions, as is the case with spinfoams, another problem arises. Boundary data on a spacelike hypersurface implies a simultaneous measurement of fields within a codimension-1 chunk of space. This contradicts the holographic principle, which should only allow a simultaneous measurement on a codimension-2 surface. Again, this argument doesn't invalidate spinfoam-like models, because the codimension-1 boundary data is merely ``kinematical''. Upon solving the theory, the ``physical'' state space may turn out to be much smaller, in accord with the holographic bound. However, this again means (at best) that the formalism is far removed from the reality it describes.

One may try to avoid this conflict with holography by switching to timelike boundaries. In that case, measuring the boundary amounts to a \emph{sequence} of simultaneous measurements, each one only capturing a codimension-2 slice. However, timelike boundaries suffer from a different problem: they are not causally closed. This problem is perhaps less severe than the previous ones, and arises already in classical field theory. On a timelike boundary, one can never fully predict future measurements from a set of previous ones: the next slice to be measured is always subject to new outside influences. As a result, the configuration variables on each new slice are arbitrary. From these and from earlier measurements, the momentum variables can then be deduced. This situation contrasts with the more standard case of a spacelike boundary. There, the phase-space data on the boundary's ``initial half'' can be used to fully predict the data on the ``final half''.

To summarize, we've outlined two problems with quantum gravity in finite spacelike-bounded regions: the transverse diffeomorphism redundancy and an apparent dimensional clash with holography. Timelike boundaries suffer too from the diffeomorphism issue, and also from a lack of full predictivity (which bears no specific relation to gravity). It's worth noting how the latter two problems are avoided in AdS/CFT, where a timelike asymptotic boundary is used. First, transverse diffeomorphisms simplify at infinity, becoming conformal transformations of the boundary. These can be easily incorporated as a redundancy of a quantum theory. Second, the boundary configuration variables, which cannot be predicted from past measurements, assume the role of \emph{background} parameters, since it costs infinite energy to change them. The momentum degrees of freedom, which remain dynamical, can then be predicted from their values on an initial slice.

\subsection{Null boundaries to avoid the secondary problems} \label{sec:why_null:null}

We will now demonstrate that, at least conceptually, the ``secondary'' problems from section \ref{sec:why_null:secondary} can be avoided by using null boundaries. We stress that an actual model of this kind does not exist, and it's not clear what it would look like, even in the context of classical field theory. That question is explored on a preliminary level in sections \ref{sec:class}-\ref{sec:gravity}. The nature of the quantum theory's ``output'' is briefly discussed in section \ref{sec:discuss}.

\begin{figure}%
\centering%
\includegraphics[scale=0.75]{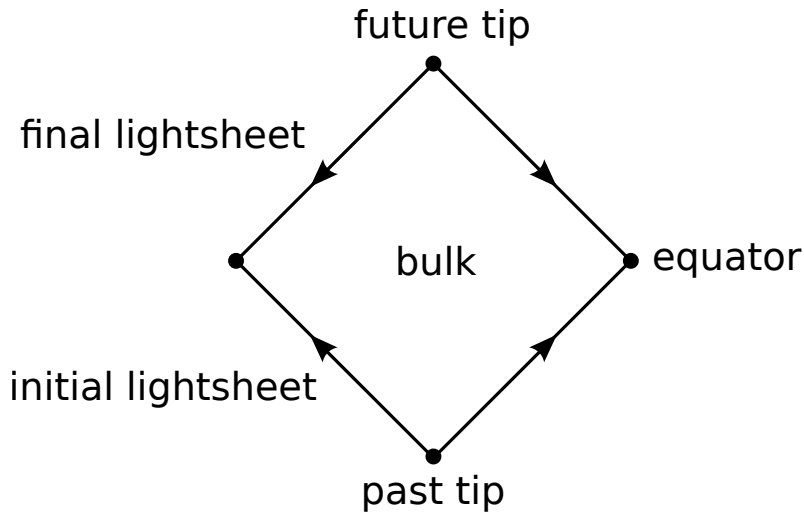} \\
\caption{Some terminology for null-bounded regions. The lines represent codimension-1 lightsheets; the dots represent codimension-2 spacelike surfaces. The arrows indicate the ``outgoing'' null normal that corresponds to the outgoing covector, for a mostly-plus metric signature. In $d>2$ dimensions with trivial topology (but not in the simple examples in section \ref{sec:class}), the left and right sides of the figure are understood to be connected.}
\label{fig:boundary} 
\end{figure}

We begin by describing the structure of null-bounded regions. Consider a spacetime metric not too different from Minkowski space. A closed null boundary is then uniquely determined by a closed codimension-2 spacelike surface. We refer to this surface as the boundary's ``equator''. The surface has precisely two lightrays orthogonal to it at each point. These two rays generate the lightcone in the 1+1d transverse timelike plane. At each point of the equator, we choose the future-pointing half of one ray and the past-pointing half of the other. Specifically, we choose the ``ingoing'' half-rays, i.e. those along which the area decreases, and which eventually intersect. Thus, we limit ourselves to ``normal'' equators in the sense of \cite{Bousso:1999xy}, and do not consider trapped or anti-trapped ones. We terminate each ray upon its first intersection with another ray. The null boundary is then given by the union of these rays. It consists of two smooth null hypersurfaces, or ``lightsheets''. Generically, the lightray intersections away from the equator constitute a pair of codimension-2 spacelike surfaces - one for the past-ingoing rays, and one for the future-ingoing ones. We refer to these surfaces as the past and future ``tips'', respectively. The boundary consists of an initial and a final half. The initial (final) half-boundary stretches from the equator to the past (future) tip. See figure \ref{fig:boundary}. If the metric is a solution of GR with suitable energy conditions, then the boundary area decreases monotonously from the equator towards the tips \cite{Bousso:1999xy}. The past and future tips are then anti-trapped and trapped surfaces, respectively. In the degenerate case where the past and future tips are single points, the two half-boundaries become lightcones. The full boundary is then a causal diamond \cite{Bousso:2000nf}.

The above recipe differs from that in \cite{Bousso:1999xy}, in that we terminate each lightray upon its first intersection. In \cite{Bousso:1999xy}, the rays are continued until they reach zero-area caustics - surfaces where they intersect with \emph{neighboring} rays. This choice is not suitable for us, since it doesn't result in a closed hypersurface. Furthermore, we will see in section \ref{sec:gravity} that nonvanishing tip areas lead to an entropy-like imaginary contribution to the GR action. The tips and the caustics coincide at the tip surfaces' edges.

With sufficient curvature or a non-trivial topology, existence and uniqueness issues arise. For instance, in a de-Sitter universe, equators above a certain size do not generate closed boundaries, because the lightrays never intersect. Another possibility is a disjoint equator. For a simple example, consider a spacetime of the form $\mathbb{R}^{1,1}\times\mathcal{M}$, with $\mathcal{M}$ closed and spacelike. In this spacetime, consider the boundary $D_1\times\mathcal{M}$, where $D_1$ is a 1+1d causal diamond. The ``equator'' of $D_1$ consists of two points, so the equator of $D_1\times\mathcal{M}$ consists of two copies of $\mathcal{M}$. The rays from each connected equator component to each tip generate a smooth lightsheet. In this case, then, the closed null boundary consists of four separate lightsheets.

We can now see how the null boundary structure avoids all the difficulties from section \ref{sec:why_null:secondary}. First, there is no extra degree of freedom allowing for transverse displacements of the boundary. In a given spacetime, the entire boundary is fixed by its codimension-2 equator. Indeed, any lightsheet consists of null geodesics, and is fully determined by a codimension-2 slice, up to a discrete choice for the rays' initial direction. 

Another way to view the disappearance of the redundancy is that it gets swallowed by the constraint equation $n_\mu n^\mu = 0$ that requires the boundary normal to be null. Yet another intuition is that the Hamiltonian constraint refers to displacements of the boundary \emph{in perpendicular} to itself. Now, on a lightsheet, the normal direction is also tangent. Thus, perpendicular displacements become intrinsic diffeomorphisms, which do not alter the boundary's location in spacetime. This implies that null boundaries are a more subtle and natural way to remove the redundancy than a conventional gauge choice.

We note a peculiar difference between the null-boundary setup and canonical GR. In the latter, the Hamiltonian constraint plays a dual role: it encodes the theory's dynamics, as well as the redundancy under transverse displacements. With null boundaries, the redundancy is automatically taken care of, so the dynamics must be provided separately, e.g. in path-integral form. This sort of ``splitting'' of an object with dual roles into two separate objects is a recurring feature in geometry and field theory on lightsheets. 

The other problems from section \ref{sec:why_null:secondary} evaporate similarly. As in the timelike case, the preparation and measurement of a null boundary state proceed in a sequence of codimension-2 snapshots, never requiring an instantaneous codimension-1 measurement. In this sense, we can avoid the clash with holography that afflicts spacelike boundaries. On the other hand, as in the spacelike case, a null boundary is causally closed: full knowledge of the fields on the initial half is sufficient to deduce their values on the final half. In this way, the predictivity issue associated with timelike boundaries is also avoided.

The restriction to null boundaries is holographic in yet another sense, closely related to the demise of the Hamiltonian constraint. In ordinary field theory, once we refuse the non-predictivity of timelike boundaries, the state of a finite region can be described in terms of any spacelike hypersurface that terminates on the same codimension-2 surface. The ``region'' is thus effectively \emph{defined} by its closed codimension-2 boundary. The limiting case of these spacelike hypersurfaces is just the null boundary defined above, with the codimension-2 surface as its equator. The null boundary is thus the \emph{outermost} boundary of a causally closed region in spacetime. By restricting to it, we are saying in some sense that the region can only be discussed ``from the outside''. This statement has a holographic ring to it, though of course it shouldn't be confused with a reduction in the degrees of freedom. It also releases us from providing a unitary description of intermediate measurements, such as spacelike intermediate states. This may relieve some of the difficulties concerning black holes and information (see the recent discussion \cite{Almheiri:2012rt,Bousso:2012as,Susskind:2012uw}). 

\subsection{The case for discrete geometries} \label{sec:why_null:discrete}

In the above, we advocated the use of null boundaries for finite-region QG. We would now like to go further, and argue that this idea might be best implemented in a discrete model. 

First, a closed null boundary cannot be everywhere smooth. Like any closed hypersurface of fixed causal type in spacetime, it must contain codimension-2 ``corners'', where the boundary is not differentiable. In fact, one may argue that even smooth closed boundaries contain degenerate surfaces in the \emph{metric} sense - the loci of signature flips. In the null case, the corner surfaces are the lightsheet (self-)intersections at the equators and tips. Now, if such singular objects are inevitable in our description of the geometry, then it's better if they are the rule, not the exception. In this way, we are led to a picture in which the building blocks of spacetime are codimension-1 lightsheets intersecting at codimension-2 surfaces. This is a natural situation for discrete models, particularly for spinfoams.

To take this point further, it's desirable for bulk histories to be described in the same terms as the boundary data. This allows an interpretation of bulk histories as collections of on-shell states. For QG with boundary data on a set of intersecting lightsheets, it is therefore natural to define bulk geometries also in terms of intersecting lightsheets. This again implies a discretization of the bulk spacetime region. We emphasize that bulk histories are to be viewed as \emph{collections}, not \emph{sequences} of on-shell states. We are not proposing to view ``jagged'' null slices as intermediate states of the full region - that would be tantamount to allowing spacelike states back in. Instead, we should view closed null boundaries within the bulk as on-shell states for the \emph{sub}regions defined by their equators. 

The discretization we are suggesting here must differ in nature from that in e.g. Regge calculus \cite{Regge:1961px}. There, spacetime is constructed out of \emph{flat} hyperfaces and their intersections. In contrast, using flat lightsheets, one cannot compose even a simple shape in Minkowski space, such as a lightcone. In particular, flat lightsheets cannot accommodate area expansion or contraction. Thus, the discrete lightsheet elements must be allowed some rudimentary form of curvature. For instance, one may specify an initial and final area for each element, allowing the two to differ.  

Constructing bulk geometries out of intersecting lightsheets makes sense also from a different perspective. Specifically, it seems to offer a compelling picture for the role of black holes. Consider a sum over bulk geometries within some null-bounded experimental region. Generically, these geometries will contain black holes, which are created and then evaporated through Hawking radiation. At least semiclassically, these black holes contain singularities. Due to evaporation, the singularities are not eternal. Thus, there exist finite ``hidden'' regions in the bulk, within which all causal lines fall into the singularity, failing to emerge outside. The boundary of each hidden region is a causal horizon, i.e. a null hypersurface. It is defined similarly to an event horizon, but with asymptotic infinity replaced by the boundary of the experimental region. Now, the quantum field fluctuations within the hidden regions should not affect any outside observables. In other words, they shouldn't contribute to the path integral. We are thus led to postulate that each hidden region should enter the sum-over-histories as a single discrete piece, with no resolution of its interior. Favoring rules over exceptions, we arrive again at a picture of bulk geometries discretized into null-bounded cells. Having come this far, we can also reverse the logic, and interpret the interior of \emph{every} discrete cell as a short-lived black hole. This picture offers a neat realization of the notion that spacetime discreteness is due to horizons at the Planck scale.

In the above discussion, the null boundary of every bulk cell is \emph{closed}: it never falls into the singularity, which lies entirely on the inside (with the possible exception of its edge). For consistency, we should demand the same for the outer boundary of the experimental region. In other words, the outer boundary should be a closed null hypersurface that never hits a singularity. It is tempting to relate this to our insistence on non-trapped equators. However, we are not aware of general results that relate the two notions. A possible independent way to enforce closure is to spread the boundary data throughout the lightsheets, which would imply that they extend as far as they should. This contrasts with the usual practice for null boundaries, which is to place all the boundary data on either the initial or the final half. Note that there may well be a singularity \emph{outside} the experimental region. In particular, the region may lie entirely or partially inside the horizon of a large black hole. All we require is that before they hit a singularity, the boundary lightsheets must ``close'' by intersecting.  

The picture sketched above resembles a form of black hole complementarity \cite{Stephens:1993an,Susskind:1993if}. A black hole's interior can affect an observation, but only if the observer himself (here, the boundary of the experimental region) falls inside. From outside a black hole, only the horizon is relevant, with no resolution of the interior. Note also that singularities never explicitly appear. They are always either outside the experiment, or within hidden regions. A boundary that hits a singularity corresponds to an experiment that couldn't be completed, and therefore doesn't need to be described.

\section{Classical field theory with null boundary data} \label{sec:class}

\subsection{Framework}

In this section, we explore the use of null finite boundaries in the context of classical field theory. As explained in the Introduction, we consider the on-shell action as the ``observable'' from which all physics is to be derived. Thus, we'll be discussing on-shell action functionals with null boundary data. The goal is to see which features of field theory can be recovered from such actions alone. In effect, we are doing classical field theory with our hands tied behind our back. The significance of this exercise is that in QG, we may be truly restricted to null-boundary observables, as argued in section \ref{sec:why_null}.

As we will see, the action functional for null-bounded regions has a number of nonstandard properties. These can all be related to a breakdown of the usual notion of canonical conjugates. The underlying reason for that is the ``halving'' of the boundary degrees of freedom that takes place in the lightlike case.

In relativistic field theory, lightsheets are characteristic hypersurfaces of the field equations. As a result, when providing boundary data on a lightsheet, we need half as many variables per point as we would on a non-null boundary (assuming all redundancies are accounted for) \cite{Penrose:1985jw}. To understand this heuristically, we note that the conjugate momentum of a field $\varphi$ is usually related to its derivative $n^\mu\del_\mu\varphi$ along the boundary's normal. But the normal to a lightsheet is also tangent to it. Therefore, the value of $\varphi$ on the lightsheet automatically contains the derivative $n^\mu\del_\mu\varphi$. The momentum data is thus included in the configuration data, which explains the ``halving'' of the degrees of freedom. 

The above already suggests that something unusual should happen to the notion of canonical conjugates. The point can be sharpened, as follows. Due to the ``halved'' degrees of freedom, a classical solution is now determined by the configuration data on just the initial half-boundary. In particular, the data on the final half is determined as well. As a result, we cannot vary the configuration data at a boundary point without affecting it at other points. But such localized variations are essential for the standard properties of canonical conjugates, as derived from the action. This point can be easily missed if one proceeds formally with the Lagrangian derivation, habitually treating variations at the initial and final half-boundaries as independent. 

There is an important difference between our approach and what is normally meant by ``physics on null hypersurfaces''. Usually, one takes e.g. a formula for the charge current, and projects it onto a null hypersurface (which may or may not be at asymptotic infinity). The formula itself, however, is obtained by conventional methods of field theory, which ultimately rely on action variations with non-null boundary data. This approach can be termed ``null later'': the null nature of the boundary enters late in the conceptual chain. In contrast, our present approach is ``null first'': we wish to use only concepts that can be derived from the actions of null-bounded regions. Thus, we do not take for granted the usual constructs of field theory, even when they have a sensible null limit. The issue is whether they can be \emph{derived} from null-bounded actions alone. 

On the other hand, when calculating the null-bounded actions themselves, we \emph{will} use standard tools such as field equations, and we will take limits of non-null action formulas. This double standard is inevitable, since we do not have a complete reformulation of field theory. We arrive at the null-bounded on-shell actions by whatever means we can, then forget the means and study the actions in their own right.   

As a final remark, we will not consider Hamiltonian approaches to null-bounded field theory. The reason is that there doesn't seem to be a good notion of Hamiltonian flow. Indeed, there is no way to ``evolve'' the initial half-boundary such that it remains null and still terminates on the equator, i.e. still spans the same causal region. This of course echoes the arguments made in section \ref{sec:why_null:null}. With no way to define the dynamics in Hamiltonian terms, we must resort to the action functional.

\subsection{The free scalar action} \label{sec:class:scalar_action}

We begin by considering the simplest field theory - a free scalar field in a $d$-dimensional background spacetime. We define the metric signature as mostly-plus. We take the field $\varphi$ to be complex; the results for a real $\varphi$ are similar. The off-shell action in a spacetime region $\Omega$ reads:
\begin{align}
 S = -\int_\Omega \sqrt{-g}\left(g^{\mu\nu}\del_\mu\varphi^*\del_\nu\varphi + m^2\varphi^*\varphi\right) d^dx \ . \label{eq:scalar_S}
\end{align}
The field equation is:
\begin{align}
 \del_\mu\left(\sqrt{-g}g^{\mu\nu}\del_\nu\varphi\right) = \sqrt{-g}m^2\varphi \ . \label{eq:scalar_field_eq}
\end{align}
Integrating by parts and using the field equation, the action \eqref{eq:scalar_S} reduces on-shell to a boundary term:
\begin{align}
 S = -\frac{1}{2}\int_\Omega \del_\mu\left(\sqrt{-g} g^{\mu\nu}(\varphi^*\del_\nu\varphi + \varphi\del_\nu\varphi^*)\right) d^dx 
  = -\frac{1}{2}\int_{\del\Omega} n_\mu \sqrt{-g} g^{\mu\nu}\del_\nu(\varphi^*\varphi) d^{d-1}x \ . \label{eq:scalar_S_boundary}
\end{align}
Here, $n_\mu = \epsilon_{\mu ab\dots c}\epsilon^{ab\dots c}/(d-1)!$ is the (appropriately densitized) covector associated with the codimension-1 boundary $\del\Omega$. The indices $(a,b,\dots)$ span the boundary's tangent space. 

In the null case, the codimension-1 vector density $s^\nu = n_\mu \sqrt{-g} g^{\mu\nu}$ is tangent to the boundary, and is directed along its generating lightrays. It can therefore be written with a boundary-tangent index, as $s^a$. In fact, $s^a$ is the ``area current'' of the boundary lightsheet - the null analogue of a volume density. Since the sign of $n_\mu$ in \eqref{eq:scalar_S_boundary} must be chosen as outgoing, we get (for a mostly-plus metric) that $s^a$ is future-pointing on the initial lightsheets and past-pointing on the final ones. In other words, $s^a$ points away from tips and towards equators. See figure \ref{fig:boundary}. The on-shell action \eqref{eq:scalar_S_boundary} now reads:
\begin{align}
 S = -\frac{1}{2}\int_{\del\Omega} s^a \del_a\left|\varphi\right|^2 d^{d-1}x 
  = -\frac{1}{2}\int_{\del\Omega}\left(\del_a(\left|\varphi\right|^2 s^a) - \left|\varphi\right|^2 \del_a s^a\right) d^{d-1}x \ . \label{eq:scalar_S_null}
\end{align}

The first term in \eqref{eq:scalar_S_null} can be integrated, giving the fluxes of $\left|\varphi\right|^2 s^a$ through the codimension-2 lightsheet intersections. Up to a sign, the flux of $s^a$ through any surface is the surface area density $\sqrt{\gamma}$, where $\gamma_{ij}$ is the induced codimension-2 spacelike metric. The sign is determined by the orientation of $s^a$. From the discussion above, it is positive for equator surfaces and negative for tips. Each surface contributes to the integral \emph{twice} - once for every lightsheet that ends on it. Our on-shell action finally becomes:
\begin{align}
 S = \frac{1}{2}\int_{\del\Omega} \left|\varphi\right|^2 \del_a s^a d^{d-1}x 
  - \left(\sum_{\mathrm{equators}} - \sum_{\mathrm{tips}}\right)\int\left|\varphi\right|^2\sqrt{\gamma}\, d^{d-2}x \ . \label{eq:scalar_S_final}
\end{align}
The divergence $\del_a s^a$ in the first term is the lightsheet's area expansion rate. 

Formula \eqref{eq:scalar_S_final} is remarkably simple. It is a boundary integral of $\left|\varphi\right|^2$, with an algebraic kernel determined by the geometry. This simplicity is somewhat deceptive, since the values of $\varphi$ on the entire boundary are not independent. The values on the final half-boundary are (linear) functionals of the values on the initial half. This makes the action \eqref{eq:scalar_S_final} \emph{non-local}, as we will now discuss.

\subsection{Non-locality} \label{sec:class:scalar_locality}

Let us briefly review what is meant by ``locality'' of an on-shell action. Viewed purely as a functional of boundary configuration data, the action $S$ is in fact typically non-local. A contribution that is local in the boundary data can usually be discarded, as it affects neither the dynamics nor the Poisson brackets. At first sight, this would suggest that our free-scalar action \eqref{eq:scalar_S_final} is trivial, and can be set to zero. However, in the null setup, if our goal is e.g. to derive canonical momenta from $S$, terms of this type cannot be discarded. The reason is that, as mentioned above, the boundary configuration fields are not independent in the null setup. The action \eqref{eq:scalar_S_final} is a non-trivial functional of the \emph{independent} boundary data, such as the values of $\varphi$ on the initial half-boundary.

If $S$ is usually a non-local functional of boundary data, how does it encode the locality of the field theory? The answer is that canonical momenta, i.e. the variations of $S$ with respect to boundary fields, are local \emph{in terms of the bulk solution}: they can be determined from the fields and their bulk derivatives at the variation point. It is this notion of locality that is violated in the null setup. We will now clarify this statement through examples.

Consider the simplest of free scalar theories: a massless scalar field in 1+1d Minkowski space. We choose null coordinates $(u,v)$, so that $ds^2 = -2dudv$. The field equation \eqref{eq:scalar_field_eq} reads simply $\del_u\del_v\varphi = 0$. Without loss of generality, we can choose our null-bounded region as the diamond $0\leq u,v\leq L$ (Lorentz boosts preserve only the area $L^2$). The solution within the diamond is fully determined by the boundary data on the initial edges:
\begin{align}
 \left\{\varphi(u,0),\varphi(0,v)\right\} \equiv \left\{f(u),g(v)\right\} \ , \label{eq:scalar_data}
\end{align}
with the constraint $f(0) = g(0)$. The solution for this boundary data reads:
\begin{align}
 \varphi(u,v) = f(u) + g(v) - f(0) \ . \label{eq:scalar_massless_solution}
\end{align}

Let us evaluate the action \eqref{eq:scalar_S_final} for this setup. In 1+1d, the ``area expansion rate'' vanishes identically, so the first term doesn't contribute. The past and future ``tip surfaces'' are the points $(0,0)$ and $(L,L)$, respectively. The ``equator surfaces'' are the two points $(L,0)$ and $(0,L)$. The ``area density'' $\sqrt{\gamma}$ and the codimension-2 integrals are trivial. We get:
\begin{align}
 S = \left|\varphi(L,L)\right|^2 + \left|\varphi(0,0)\right|^2 - \left|\varphi(L,0)\right|^2 - \left|\varphi(0,L)\right|^2 \ . \label{eq:scalar_2d_action}
\end{align}
On-shell, the value $\varphi(L,L)$ at the future tip is a function of the $\varphi$ values at the other three corners:
\begin{align}
 \varphi(L,L) = \varphi(L,0) + \varphi(0,L) - \varphi(0,0) \ .
\end{align}
Using this relation, the action can be expressed in terms of $\left\{\varphi(0,0),\varphi(L,0),\varphi(0,L)\right\}$ alone:
\begin{align}
 \begin{split}
   S &= \left(\varphi(L,0) - \varphi(0,0)\right)\left(\varphi^*(0,L) - \varphi^*(0,0)\right) + \mathrm{c.c.} \\
    &= \left(f(L) - f(0)\right)\left(g^*(L) - g^*(0)\right) + \mathrm{c.c.}
 \end{split} \label{eq:scalar_2d_massless_action}
\end{align}
This expression for $S$ in terms of independent boundary data is non-local, in the following sense. Consider e.g. the derivative of $S$ with respect to the boundary value $\varphi(L,0)$:
\begin{align}
 \frac{\delta S}{\delta\varphi(L,0)} = \varphi^*(0,L) - \varphi^*(0,0) \ . \label{eq:scalar_2d_massless_variation}
\end{align}
This derivative involves field values far from the point $(L,0)$. As discussed above, it's not enough that \eqref{eq:scalar_2d_massless_variation} depends on \emph{boundary} data away from $(L,0)$. The non-locality statement rests on the fact that the ``momentum'' \eqref{eq:scalar_2d_massless_variation} cannot be determined from the field \emph{in a small bulk neighborhood} of the variation point. Indeed, the solution \eqref{eq:scalar_massless_solution} in the bulk neighborhood of $(L,0)$ can only tell us about $f(u)$ around $u=L$ and $g(v)$ around $v=0$, as well as their complex conjugates. On the other hand, the ``momentum'' \eqref{eq:scalar_2d_massless_variation} depends on $g^*(v)$ at $v=L$. 

A second example will allow us to sharpen this notion of non-locality. Consider a \emph{massive} free scalar in 1+1d. The boundary data can be chosen as before. The field equation \eqref{eq:scalar_field_eq} reads:
\begin{align}
 \del_u\del_v\varphi = -\frac{m^2}{2}\varphi \ .
\end{align}
This can be solved recursively, order by order in $m^2$. The solution reads:
\begin{align}
 \begin{split}
   \varphi(u,v) ={}& f(u) + g(v) - \sum_{n=0}^\infty \frac{(-m^2 uv/2)^n}{(n!)^2}f(0) \\
    &+ \sum_{n=1}^\infty \frac{(-m^2/2)^n}{n!(n-1)!}\left(v^n\int_0^u {(u-u')^{n-1} f(u')du'} + u^n\int_0^v {(v-v')^{n-1} g(v')dv'} \right) \ .
 \end{split}
\end{align}
The expression \eqref{eq:scalar_2d_action} for the on-shell action still holds. This time, since information doesn't travel only at the speed of light, the value $\varphi(L,L)$ at the future tip cannot be deduced from the $\varphi$ values at the other corners. Thus, the four field values appearing in \eqref{eq:scalar_2d_action} are independent. At first sight, this means that the action is a local function of independent boundary data. This would imply that the non-locality of $S$ is peculiar to the massless case. However, a slightly refined notion of non-locality \emph{does} hold for the massive scalar, and in fact seems to be generic. To formulate this notion, we will need the concept of a \emph{local set} of boundary data. We define this as a collection of boundary data with connected support that determines a unique bulk solution. The data set \eqref{eq:scalar_data} is a valid example. However, the field values at the four corners, though independent, are not part of any local set. Indeed, such a set would have to involve at least three of the diamond's edges, and the $\varphi$ values on three edges overspecify the solution. If we insist on a local data set such as $\left\{f(u),g(v)\right\}$, then the field value $\varphi(L,L)$ in \eqref{eq:scalar_2d_action} becomes a non-trivial function of the boundary data:
\begin{align}
 \begin{split}
   \varphi(L,L) ={}& f(L) + g(L) - \sum_{n=0}^\infty \frac{(-m^2 L^2/2)^n}{(n!)^2}f(0) \\
    &+ \sum_{n=1}^\infty \frac{(-m^2 L/2)^n}{n!(n-1)!}\left(\int_0^L {(L-u)^{n-1} f(u)du} + \int_0^L {(L-v)^{n-1} g(v)dv} \right) \ .
 \end{split} \label{eq:phi_LL}
\end{align}
Plugging this into \eqref{eq:scalar_2d_action}, we get the action in terms of $\left\{f(u),g(v)\right\}$. Its variation with respect to e.g. a value $f(u)$ with $u\neq 0,L$ reads:
\begin{align}
 \frac{\delta S}{\delta f(u)} = \varphi^*(L,L)\frac{\delta\varphi(L,L)}{\delta f(u)}
  = \varphi^*(L,L)\sum_{n=1}^\infty \left(-\frac{m^2 L}{2}\right)^n \frac{(L-u)^{n-1}}{n!(n-1)!} \ . \label{eq:scalar_2d_massive_variation}
\end{align} 
The factor $\delta\varphi(L,L)/\delta f(u)$ is a field-independent function of the coordinate $u$. On the other hand, the factor $\varphi^*(L,L)$, given by the complex conjugate of \eqref{eq:phi_LL}, depends on the boundary data $\left\{f^*(u'),g^*(v')\right\}$ in the entire range $0<u',v'<L$. In particular, it depends on field values far from the variation point $(u,0)$. Again, the non-trivial statement is that $\delta S/\delta f$ can't be determined from a small \emph{bulk} neighborhood of $(u,0)$. For instance, if the variation point is near the past tip, $\delta S/\delta f$ still depends on $\left\{f^*(u'),g^*(v')\right\}$ values near the equator, which cannot be determined from the tip's neighborhood.

To summarize, the on-shell action in a null-bounded region is non-local in the following sense: there is no local (i.e. connectedly supported) set of independent boundary data, such that the action variation with respect to it can be determined from the bulk solution in a small neighborhood of the variation point.

\subsection{Explicit and local-set ``holography''} \label{sec:class:scalar_holography}

Consider again the expression \eqref{eq:scalar_S_final} for the null-bounded on-shell action of a free scalar. When the boundary's geometry is such that the expansion rate $\del_a s^a$ vanishes, the action is reduced to codimension-2 integrals at the lightsheet intersections. For instance, in the 1+1d cases studied above, the codimension-2 ``surfaces'' are the diamond's corners, and the action is determined by the field values on them, as in \eqref{eq:scalar_2d_action}. If one believes that all physics should be contained in the on-shell action, then this looks like a holographic reduction in the degrees of freedom. We will refer to this property as ``explicit holography''. Unlike the non-locality discussed in section \ref{sec:class:scalar_locality}, this isn't a general property of null boundary data. As far as we can tell, it only applies to free scalars with $\del_a s^a = 0$ and to pure gravity (see section \ref{sec:gravity}).

There is another sense in which the free scalar action with $\del_a s^a = 0$ is ``holographic''. Recall that the codimension-2 boundary data on which the second term in \eqref{eq:scalar_S_final} depends isn't part of a local data set. One can ask whether some form of ``holography'' would still take place in terms of a local set. For the massless free 1+1d scalar, the answer is clearly positive. Indeed, the action \eqref{eq:scalar_2d_massless_action} is written in terms of the local boundary data \eqref{eq:scalar_data}, and still depends only on field values at isolated points, i.e. at codimension-2 ``surfaces''. However, this doesn't seem to happen in any other setup. For instance, eqs. \eqref{eq:scalar_2d_action} and \eqref{eq:phi_LL} show that for a massive scalar, the action depends on the entire codimension-1 local set $\left\{f(u),g(v)\right\}$. But if we vary the action with respect to this local set, as in \eqref{eq:scalar_2d_massive_variation}, we find a codimension-\emph{two} set of ``canonical momenta''. Indeed, \eqref{eq:scalar_2d_massive_variation} tells us that the ``momenta'' $\delta S/\delta f(u)$ at different $u$'s are all given by the same functional $\varphi^*(L,L)$, up to field-independent coefficients $\delta\varphi(L,L)/\delta f(u)$. This is in fact a general feature of free scalar actions with $\del_a s^a = 0$. To see this, note that the value of $\varphi$ at each point is a linear functional of the boundary data. The second term in \eqref{eq:scalar_S_final} is a quadratic form over a codimension-2 set of such functionals. It follows that this same set of functionals spans all the derivatives of $S$. 

If the action variations with respect to independent local boundary data are spanned by a codimension-2 set of ``canonical momenta'', we will say that the action exhibits ``local-set holography''. We saw that for free scalars with $\del_a s^a = 0$, this property follows from ``explicit holography'' and from the linearity of the field equations. In section \ref{sec:class:maxwell}, we will show that the ``explicit holography'' is not a necessary condition. It seems likely that linearity \emph{is} necessary, but we will not attempt to prove this.  

So far, we've explicitly calculated $S$ in terms of local data sets only for 1+1d scalars. In higher dimensions, the boundary's expansion rate no longer vanishes automatically. In particular, the ``holographic'' properties described above are not expected to hold for a causal diamond in $d>2$ dimensions. When we do have a higher-dimensional boundary with $\del_a s^a = 0$, the free scalar theory (whether massless or massive) is qualitatively similar to the 1+1d massive case. Regardless of mass, propagation along codimension-2 spacelike surfaces makes the motion in the 1+1d transverse plane slower than light. The simplest case is a flat spacetime with topology $\mathbb{R}^{1,1}\times T^{d-2}$, where $T^{d-2}$ is a torus. We can then choose a boundary of the form $D_1\times T^{d-2}$, where $D_1$ is a 1+1d causal diamond. Using null coordinates $(u,v)$ in $\mathbb{R}^{1,1}$, we Fourier-transform the fields along the compact coordinates $x^i$:
\begin{align}
 \varphi(u,v,x^i) = \frac{1}{\sqrt{V_{d-2}}}\sum_\boldk \varphi_\boldk(u,v) e^{ik_i x^i} \ ,
\end{align}
where $V_{d-2}$ is the area of the codimension-2 torus. From here on, the results are completely analogous to those in section \ref{sec:class:scalar_locality}. The field equation reads:
\begin{align}
 \del_u\del_v\varphi_\boldk(u,v) = -\frac{m^2 + k^2}{2}\varphi_\boldk(u,v) \ .
\end{align}
The on-shell action \eqref{eq:scalar_S_final} reads:
\begin{align}
 S = \sum_\boldk \left(\left|\varphi_\boldk(L,L)\right|^2 + \left|\varphi_\boldk(0,0)\right|^2 - \left|\varphi_\boldk(L,0)\right|^2 - \left|\varphi_\boldk(0,L)\right|^2\right) \ .
 \label{eq:scalar_torus_action}
\end{align}
A local set of boundary data is given by:
\begin{align}
 \left\{\varphi_\boldk(u,0),\varphi_\boldk(0,v)\right\} \equiv \left\{f_\boldk(u),g_\boldk(v)\right\} \ , \label{eq:scalar_data_k} 
\end{align}
with $f_\boldk(0) = g_\boldk(0)$. The future-tip values $\varphi_\boldk(L,L)$ are expressed in terms of the local set as:
\begin{align}
 \begin{split}
   \varphi_\boldk&(L,L) = f_\boldk(L) + g_\boldk(L) - \sum_{n=0}^\infty \frac{(-(m^2 + k^2) L^2/2)^n}{(n!)^2}f_\boldk(0) \\
    &+ \sum_{n=1}^\infty \frac{(-(m^2 + k^2)L/2)^n}{n!(n-1)!}
     \left(\int_0^L {(L-u)^{n-1} f_\boldk(u)du} + \int_0^L {(L-v)^{n-1} g_\boldk(v)dv} \right) \ .
 \end{split}
\end{align}
As before, the variations of $S$ with respect to the local set \eqref{eq:scalar_data_k} are spanned by a codimension-2 set of ``canonical momenta''. These consist of the boundary data $\varphi_\boldk(0,0),\varphi_\boldk(L,0),\varphi_\boldk(0,L)$ and the linear functional $\varphi_\boldk(L,L)$, together with their complex conjugates, for each codimension-2 momentum $\boldk$. 

A non-``holographic'' example can be easily constructed by perturbing the $D_1\times T^{d-2}$ boundary shape. Consider a small constant expansion rate $\del_a s^a = \alpha$ on the initial half-boundary, keeping $\del_a s^a = 0$ on the final half. To accommodate this, the bulk metric should be modified by varying the $T^{d-2}$ volume as a function of position in $\mathbb{R}^{1,1}$. ``Explicit holography'' is then trivially lost, since the first term in \eqref{eq:scalar_S_final} no longer vanishes. Specifically, it is given by $\alpha$ times the integral of $\left|\varphi\right|^2$ over the initial half-boundary. Now, the values of $\varphi$ on the initial half-boundary are a local data set. Thus, the first term in \eqref{eq:scalar_S_final} is a non-degenerate quadratic form over a codimension-1 local set. The second term is a quadratic form of codimension-2 ``rank'', by the same arguments as before. The ``rank'' of the overall action is therefore dominated by the first term. As before, this ``rank'' is also the ``number'' of independent canonical momenta. We conclude that there is a codimension-1 set of independent momenta. Thus, ``local-set holography'' is also absent.

\subsection{Free Maxwell field} \label{sec:class:maxwell}

Let us now turn to the on-shell null-bounded action of free Maxwell theory in a background spacetime. In the simplest 2+1d geometry, we will find that ``local-set holography'' is satisfied, but ``explicit holography'' is not. This implies that ``explicit holography'' is a special property of free scalars (together with pure gravity - see section \ref{sec:gravity}). Our derivation of ``local-set holography'' makes use of the spacetime dimension. It may be overturned in 3+1d or higher, as well as in more complicated 2+1d geometries. As with the scalar theory, we will also find that the action is non-local, in the sense of section \ref{sec:class:scalar_locality}. Though we will demonstrate this only in the 2+1d calculation, we expect this property to hold generally.

A free Maxwell field has the following action and field equations:
\begin{align}
 & S = -\frac{1}{4}\int_\Omega \sqrt{-g}F_{\mu\nu}F^{\mu\nu} d^dx; \label{eq:max_action} \\
 & \del_\mu\left(\sqrt{-g}F^{\mu\nu}\right) = 0;\quad \del_{[\mu} F_{\nu\rho]} = 0 \ . \label{eq:max_equations}
\end{align}
The field strength $F_{\mu\nu}$ can be expressed in terms of a gauge potential as $F_{\mu\nu} = 2\del_{[\mu}A_{\nu]}$. On-shell, the action \eqref{eq:max_action} can then be written as a boundary term:
\begin{align}
 \begin{split}
   S &= -\frac{1}{2}\int_\Omega \sqrt{-g}\del_\mu A_\nu F^{\mu\nu} d^dx = -\frac{1}{2}\int_\Omega\del_\mu\left(\sqrt{-g} A_\nu F^{\mu\nu}\right) d^dx \\
    &= -\frac{1}{2}\int_{\del\Omega}n_\mu \sqrt{-g} A_\nu F^{\mu\nu} d^dx = -\frac{1}{2}\int_{\del\Omega}s^\mu A^\nu F_{\mu\nu} d^dx \ ,
 \end{split} \label{eq:max_shell_action}
\end{align}
where we again used the area current $s^\mu = \sqrt{-g}g^{\mu\nu}n_\nu$, oriented as in section \ref{sec:class:scalar_action}. In the free scalar theory, for special boundary shapes, we could further reduce the action to a sum of codimension-2 integrals. This cannot be done with the Maxwell action \eqref{eq:max_shell_action}, except in the degenerate 1+1d case. Thus, in the language of section \ref{sec:class:scalar_holography}, free Maxwell theory is not ``explicitly holographic''. 

We now turn to examine the simplest lower-dimensional setups. In 1+1d, the theory is trivial, with no dynamical degrees of freedom. In coordinates $(u,v)$, the field equations set $\sqrt{-g}F^{vu} \equiv F$ to a constant. The action \eqref{eq:max_action} is simply $F^2/2$ times the spacetime volume of the region. Needless to say, the action's dependence on just a single field value $F$ should not be confused with the ``holography'' exhibited by free scalars.

The simplest non-trivial setup is a 2+1d flat spacetime with topology $\mathbb{R}^{1,1}\times S_1$ - a special case of the toroidal topology $\mathbb{R}^{1,1}\times T^{d-2}$. As in section \ref{sec:class:scalar_holography}, we choose a boundary of the form $D_1\times S_1$, and use the same coordinate conventions. We Fourier-transform along the compact coordinate $x$, imposing the reality condition $A^*_{\mu,k}(u,v) = A_{\mu,-k}(u,v)$. For each wavenumber $k$, the action \eqref{eq:max_shell_action} reduces to a sum over the four edges of the diamond $0\leq u,v\leq L$. The Maxwell equations \eqref{eq:max_equations} read:
\begin{align}
 \del_u F_{uv,k} &= ikF_{ux,k}; & \del_v F_{ux,k} &= \frac{ik}{2}F_{uv,k}; \nonumber \\
 \del_v F_{uv,k} &= -ikF_{vx,k}; & \del_u F_{vx,k} &= -\frac{ik}{2}F_{uv,k} \ . \label{eq:max_equations_3d}
\end{align}
A local set of boundary data is given by:
\begin{align}
 \left\{F_{uv,k}(0,0), F_{ux,k}(u,0), F_{vx,k}(0,v)\right\} \equiv \left\{F_k, f_k(u), g_k(v)\right\} \ . \label{eq:max_data}
\end{align}
Note that the two values $F_k$ and $F_{-k}$, though complex, correspond to two real data pieces, due to the reality condition $F^*_k = F_{-k}$. For $k=0$, we have a single real data piece $F_0$. The same applies to the functions $f_k(u)$ and $g_k(v)$.

To proceed, we must distinguish between the $k=0$ and $k\neq 0$ modes. For $k=0$, eqs. \eqref{eq:max_equations_3d} have the simple solution:
\begin{align}
 F_{uv,0}(u,v) = F_0; \quad F_{ux,0}(u,v) = f_0(u); \quad F_{vx,0}(u,v) = g_0(v) \ . \label{eq:max_solution_0}  
\end{align}
The contribution of this mode to the action can now be found directly from \eqref{eq:max_action}:
\begin{align}
 \begin{split}
   S_0 &= -\frac{1}{4}\int_0^L\int_0^L F_{\mu\nu,0}F^{\mu\nu}_0 du dv = \int_0^L\int_0^L \left(\frac{1}{2}F_{uv,0}^2 + F_{ux,0}F_{vx,0} \right) du dv \\
    &= \frac{1}{2}L^2 F_0^2 + \int_0^L f_0(u) du \int_0^L g_0(v) dv \ .
 \end{split} \label{eq:max_S_0}
\end{align}
Note that $F_0$, $f_0(u)$ and $g_0(v)$ each contain an implicit $x$ integral. The first term in \eqref{eq:max_S_0} is analogous to the 1+1d result. The second term is the interesting one. It should be contrasted with the free-scalar expression \eqref{eq:scalar_2d_massless_action}. By the same arguments as for the scalar theory, the second term is non-local in the $(u,v)$ plane. Both terms are consistent with ``local-set holography'', being simple products of linear functionals. However, the second term is \emph{not} consistent with ``explicit holography'': the integrals in \eqref{eq:max_S_0} cannot be reduced to contributions from the diamond's corners. Indeed, at the corners, the solution \eqref{eq:max_solution_0} is only sensitive to $f_0(u)$ and $g_0(v)$ at the edges of their argument's range. 

Let us now consider the $k\neq 0$ modes. These will be qualitatively similar to the free scalar case. For $k\neq 0$, we can choose a gauge where $A_x$ vanishes (this is impossible for $k=0$, due to the non-trivial holonomy around the circle). With this choice, we have the following relations between the potentials and the field strengths:
\begin{align}
 F_{uv,k} = \del_u A_{v,k} - \del_v A_{u,k}; \quad F_{ux,k} = -ikA_{u,k}; \quad F_{vx,k} = -ikA_{v,k} \ .
\end{align}
Lowering indices with $g_{uv} = -1$ and keeping in mind the orientation of $s^a$, the contribution to the action \eqref{eq:max_action} of each $k\neq 0$ mode becomes:
\begin{align}
 \begin{split}
   S_k ={}& {-}\frac{1}{2}\left(\int_0^L\left(A^*_{u,k}(u,L) F_{uv,k}(u,L) - A^*_{u,k}(u,0) F_{uv,k}(u,0)\right) du \right. \\
    &\left.\quad - \int_0^L\left(A^*_{v,k}(L,v) F_{uv,k}(L,v) - A^*_{v,k}(0,v) F_{uv,k}(0,v)\right) dv \right) \\
     ={}& \frac{i}{2k}\left(\int_0^L\left(F^*_{ux,k}(u,L) F_{uv,k}(u,L) - F^*_{ux,k}(u,0) F_{uv,k}(u,0)\right) du \right. \\
    &\left.\quad - \int_0^L\left(F^*_{vx,k}(L,v) F_{uv,k}(L,v) - F^*_{vx,k}(0,v) F_{uv,k}(0,v)\right) dv \right) \ ,
 \end{split} \label{eq:S_k_raw}
\end{align}
with $S_{-k} = S_k^*$. Using the field equations \eqref{eq:max_equations_3d}, we can rewrite \eqref{eq:S_k_raw} as: 
\begin{align}
 \begin{split}
   S_k ={}& {-}\frac{1}{2k^2}\left(\int_0^L\left(F_{uv,k}(u,L)\del_u F^*_{uv,k}(u,L) - F_{uv,k}(u,0)\del_u F^*_{uv,k}(u,0) \right) du \right. \\
    &\left.\quad + \int_0^L\left(F_{uv,k}(L,v)\del_v F^*_{uv,k}(L,v) - F_{uv,k}(0,v)\del_v F^*_{uv,k}(0,v) \right) dv \right) \ .
 \end{split} \label{eq:S_k}
\end{align}
Together with the complex-conjugate contribution, this reduces to a sum of codimension-2 terms:
\begin{align}
 \begin{split}
   S_k + S_{-k} ={}& {-}\frac{1}{2k^2}\left(\int_0^L\del_u\left(\left|F_{uv,k}(u,L)\right|^2 - \left|F_{uv,k}(u,0)\right|^2 \right) du \right. \\
    &\left.\quad + \int_0^L\del_v\left(\left|F_{uv,k}(L,v)\right|^2 - \left|F_{uv,k}(0,v)\right|^2 \right) dv \right) \\
   ={}& {-}\frac{1}{k^2}\left(\left|F_{uv,k}(L,L)\right|^2 + \left|F_{uv,k}(0,0)\right|^2 - \left|F_{uv,k}(0,L)\right|^2 
    - \left|F_{uv,k}(L,0)\right|^2 \right) \ .
 \end{split} \label{eq:S_k_-k} 
\end{align}
The full action is thus given by:
\begin{align}
 \begin{split}
   S ={}& \sum_k S_k = S_0 + \sum_{k>0}(S_k + S_{-k}) \\
     ={}& \frac{1}{2}L^2 F_0^2 + \int_0^L f_0(u) du \int_0^L g_0(v) dv \\
     &- \sum_{k>0}\frac{1}{k^2}\left(\left|F_{uv,k}(L,L)\right|^2 + \left|F_{uv,k}(0,0)\right|^2 - \left|F_{uv,k}(0,L)\right|^2 
      - \left|F_{uv,k}(L,0)\right|^2 \right) \ .
 \end{split} \label{eq:max_S_3d}
\end{align}
We see that the $k\neq 0$ terms are ``explicitly holographic''. From this and from the theory's linearity, it follows that they also satisfy ``local-set holography''. For completeness, we should express the $F_{uv,k}$ values in \eqref{eq:max_S_3d} in terms of the local data set \eqref{eq:max_data}. Order by order in $k$, the full solution of the field equations \eqref{eq:max_equations_3d} reads:
\begin{align}
 \begin{split}
   F_{uv,k}&(u,v) = \sum_{n=0}^\infty\frac{(-k^2 uv/2)^n}{(n!)^2} F_k \\
     &+ ik\sum_{n=0}^\infty\frac{(-k^2/2)^n}{(n!)^2}\left(v^n\int_0^u(u-u')^n f_k(u')du' - u^n\int_0^v(v-v')^n g_k(v')dv'\right) \\
   F_{ux,k}&(u,v) = f_k(u) + \frac{ikv}{2}\sum_{n=0}^\infty\frac{(-k^2 uv/2)^n}{n!(n+1)!} F_k \\
     &+ \sum_{n=1}^\infty\frac{(-k^2/2)^n}{n!(n-1)!}\left(v^n\int_0^u(u-u')^{n-1} f_k(u')du' - u^{n-1}\int_0^v(v-v')^n g_k(v')dv'\right) \\
   F_{vx,k}&(u,v) = g_k(v) - \frac{iku}{2}\sum_{n=0}^\infty\frac{(-k^2 uv/2)^n}{n!(n+1)!} F_k \\
     &- \sum_{n=1}^\infty\frac{(-k^2/2)^n}{n!(n-1)!}\left(v^{n-1}\int_0^u(u-u')^n f_k(u')du' - u^n\int_0^v(v-v')^{n-1} g_k(v')dv'\right) \ .
 \end{split}
\end{align}
We can now extract the $F_{uv,k}$ values appearing in the action \eqref{eq:max_S_3d}:
\begin{align}
 \begin{split}
   F_{uv,k}(0,0) ={}& F_k \\
   F_{uv,k}(L,0) ={}& F_k + ik\int_0^L f_k(u)du \\
   F_{uv,k}(0,L) ={}& F_k - ik\int_0^L g_k(v)dv \\
   F_{uv,k}(L,L) ={}& \sum_{n=0}^\infty\frac{(-k^2 L^2/2)^n}{(n!)^2} F_k \\
     &+ ik\sum_{n=0}^\infty\frac{(-k^2 L/2)^n}{(n!)^2}\left(\int_0^L(L-u)^n f_k(u)du - \int_0^L(L-v)^n g_k(v)dv\right) \ .
 \end{split} \label{eq:max_solution} 
\end{align}
The $k\neq 0$ terms in the action \eqref{eq:max_S_3d} are non-local in the $(u,v)$ plane, just like the $k=0$ terms. We will not dwell on this, since the non-locality of $S_0$ is enough to render the whole action non-local. 

To conclude, we have analyzed the on-shell null-bounded action of a free Maxwell field in a simplistic 2+1d geometry. This action is non-local (in both the $k=0$ and $k\neq 0$ pieces, though one is enough), exhibits ``local-set holography'' (in both the $k=0$ and $k\neq 0$ pieces), but is not ``explicitly holographic'' (due to the $k=0$ piece).  
 
\subsection{Global charges cannot be derived}

We now come to the third property of null-bounded actions advertised in the Introduction: they cannot be used to derive global conserved charges. By now, this should not come as a big surprise. Indeed, conserved charges are a special kind of canonical momenta, and we've already seen that those behave very differently in the null formalism.

To derive the global charge of some finite-region classical solution, one usually performs the associated symmetry transformation on the boundary data. The action variation in response to transforming the fields at a boundary point gives the charge density at that point. The variation from transforming the entire initial (or final) boundary data gives the total charge. Of course, the variation from transforming \emph{both} the initial and final data is zero, since this amounts to a global symmetry applied to the entire solution.

In the null setup, this recipe no longer works. The reason, as usual, is that the boundary values of configuration fields are not independent. One cannot vary them at one point without affecting other points. In particular, if we try to derive the total charge by transforming the initial boundary data, we will always get zero. This is because the initial data fully determines the solution, including the final data. A transformation of the initial data is thus a global symmetry operation, which leaves the action invariant.

There exist other methods for defining global charges, which do extend to the null case. One can perform a transformation with an $x^\mu$-dependent parameter $\alpha$ \emph{in the bulk}, and then derive the current $J^\mu$ as the coefficient of $\del_\mu\alpha$ in the off-shell action variation. Alternatively, one can turn on a small external gauge field $A_\mu$ in the bulk, and derive $J^\mu$ as the coefficient of $A_\mu$ in the on-shell action. However, we consider both methods unsatisfactory. At the level of classical field theory, the first method requires off-shell actions, while the second requires fictitious fields. Neither method derives the global charge from the on-shell action of the actual field theory in question. At the quantum level, both the off-shell and the external-field method translate into expectation values of bulk operators. In quantum gravity, which is our true object of interest, such quantities are problematic, since there is no background spacetime in which bulk operators may be localized. We conclude that in the null formalism, global charges cannot be derived in a satisfactory way.

The above arguments can be tested on the simple case of a free massless scalar in 1+1d (section \ref{sec:class:scalar_locality}). Consider a single complex scalar, with a global charge that generates phase rotations. Without our self-imposed restrictions, it's easy to find the charge current. One can just use the general formula derived from non-null boundaries (the ``null-later approach''), or rederive it using e.g. the off-shell method. Integrating this current over the initial half-boundary, we get the following expression for the charge in terms of initial data:
\begin{align}
 \begin{split}
   Q ={}& i\int_0^L \left(\varphi(u,0)\ddsimple{\varphi^*(u,0)}{u} - \varphi^*(u,0)\ddsimple{\varphi(u,0)}{u}\right) du \\ 
     &+ i\int_0^L \left(\varphi(0,v)\ddsimple{\varphi^*(0,v)}{v} - \varphi^*(0,v)\ddsimple{\varphi(0,v)}{v}\right) dv \ .
 \end{split} \label{eq:global_Q}
\end{align}
Unlike similar previous expressions, this one can't be integrated into codimension-2 terms. In other words, it genuinely depends on the values of $\varphi$ on the entire half-boundary. In contrast, the on-shell action \eqref{eq:scalar_2d_massless_action} is only sensitive to the $\varphi$ values at the corners $(0,0),(L,0),(0,L)$. We conclude that the charge \eqref{eq:global_Q} cannot be derived from the action \eqref{eq:scalar_2d_massless_action}. This conclusion doesn't assume any particular \emph{method} of derivation. The action just doesn't contain the necessary information.

\subsection{Deriving gauged charges}

While global charges are expected to disappear in QG, gauged charges are not. The crucial difference is that a gauged charge can be measured ``holographically'' via the Gauss law on a codimension-2 enclosing surface, whereas a global charge must be integrated over a codimension-1 volume slice. More precisely, we expect \emph{abelian} gauged charges to survive. A non-abelian Gauss law is problematic, since the flux cannot be integrated gauge-covariantly over an extended surface (in AdS/CFT, this problem is solved by a preferred gauge on the boundary).

We would like the ``null-first'' formalism of classical field theory to reflect the properties of QG in this context as well. Thus, we'd like to be able to derive gauged abelian charges. This appears to be possible, at least in a large class of theories that includes GR and the bosonic sector of the Standard Model Lagrangian. Though the construction proposed below ``bends the rules'' somewhat, it doesn't seem to violate the philosophy of this paper.

With non-null boundary data, a gauged charge can be derived from the action by varying the boundary value of the gauge potential. In standard Maxwell theory, the resulting action variation is the boundary value of the electric field. More generally, it is the Noether potential for the gauge symmetry \cite{WaldNoetherPotential,Jacobson:1993vj}. The charge enclosed by a codimension-2 surface on the boundary is given, via Gauss's law, by the flux of this quantity. With a null boundary, this general procedure no longer works, for the usual reason: the boundary field values cannot be varied independently.

We will now propose an alternative method to derive the gauged charge, which makes use of the boundary's null nature. It involves a sort of gauge transformation, which affects the action without perturbing the bulk solution. We take the charged fields to be scalars and denote them by $\varphi$. They may carry internal indices, which we suppress in the notation. The gauge potential $A_\mu$ enters the Lagrangian $\mathcal{L}$ through covariant derivatives $\nabla_\mu\varphi$ and through the field strength $F_{\mu\nu}$. We assume that the Lagrangian is quadratic in derivatives, so that boundary problems are well-posed. In particular, it is quadratic in the derivatives $\nabla_\mu\varphi$ of the charged fields. We assume that the second-order terms take the form $\nabla_\mu\varphi\nabla^\mu\varphi$, so that the gradient indices are contracted with each other. We also assume that linear terms in $\nabla_\mu\varphi$ are not multiplied by any other derivatives. 

Now, consider a classical solution of the field theory in a null-bounded region. Let us focus on one of the boundary lightsheets. In its neighborhood, we define a coordinate $r$ such that the lightsheet is at $r = 0$, with the bulk at $r>0$. The lightsheet coordinates are denoted by $x^a$. Let us now vary the gauge potential as follows:
\begin{align}
 \delta A_\mu = \alpha\del_\mu r \lim_{h\rightarrow 0^+}\delta(r - h) \equiv \alpha\delta_{\mathrm{in}}(r)\del_\mu r \ . \label{eq:dA}
\end{align}
Here, $\alpha$ is a constant. The ``delta function'' $\delta_{\mathrm{in}}(r)$ has a unit integral when integrated up to the boundary from the inside. In components, we have $\delta A_r = \alpha\delta_{\mathrm{in}}(r)$ and $\delta A_a = 0$. This change in $A_\mu$ effectively offsets the boundary lightsheet from the bulk by a gauge phase $\alpha$.

Let us analyze the effects of \eqref{eq:dA} on the classical solution and its action. We argue that the bulk field configuration with the variation \eqref{eq:dA} remains a solution, with no need for any compensating variations. First, notice that the only component of $\del_\mu A_\nu$ that gets altered by \eqref{eq:dA} is $\del_r A_r$. Therefore, the field strength $F_{\mu\nu} = 2\del_{[\mu}A_{\nu]}$ remains unchanged. $A_r$ itself enters the action through the kinetic terms of the charged fields. By the assumptions above, $\del\mathcal{L}/\del A_r \sim \del\mathcal{L}/\del(\nabla_r\varphi)$ is a linear function of $\nabla^r\varphi$, with derivative-free coefficients. Now, $\nabla^r\varphi$ is a derivative along the lightsheet normal, which is also tangent due to its null nature. Thus, the action variation due to \eqref{eq:dA} involves only configuration field values on the boundary itself. As a result, the stationarity condition on the off-shell action is unaffected. We conclude that the variation \eqref{eq:dA} doesn't disturb the classical solution.

As a corollary, the on-shell action variation $dS/d\alpha$ due to \eqref{eq:dA} is simply an integral of $\del\mathcal{L}/\del A_r$ over the lightsheet. But the latter is just the current component $J^r$, i.e. the flux of $J^\mu$ through the lightsheet. Adding up the variations $dS/d\alpha$ for the initial (or final) lightsheets, we get the total charge in the null-bounded region.

In a sense, our derivation can be viewed as a special case of the standard Gauss-law method. In this perspective, the transformation \eqref{eq:dA} on e.g. the final half-boundary is seen as the propagation of a change $\delta A_\mu$ to the \emph{initial} data on the equator. Now, at the equator, what looks like the transverse component $A_r$ on the final lightsheet is an \emph{intrinsic} component $A_a$ on the initial lightsheet, transverse to the equator itself. The charge is then viewed as the action variation in response to a change in $A_a$ on (and transversely to) an enveloping surface, just as in the standard Gauss law. This is of course a whimsical interpretation, since \eqref{eq:dA} is a gauge transformation, and cannot be truly said to ``propagate''.    

\section{Classical gravity with null boundary data} \label{sec:gravity}

In section \ref{sec:class}, we calculated some on-shell null-bounded actions for free field theories. We arrived at two kinds of results. Using the field equations to integrate by parts, but without solving them explicitly, we arrived at functionals such as \eqref{eq:scalar_S_final} and \eqref{eq:max_shell_action}. By explicitly solving the field equations, we then arrived at expressions such as \eqref{eq:scalar_torus_action}, giving the action in terms of a local data set.

In this section, we consider the on-shell, null-bounded action of General Relativity. It isn't feasible to solve the field equations in sufficient generality, so we won't be able to express the action in terms of a local set. However, it is possible to arrive at analogues of eqs. \eqref{eq:scalar_S_final} and \eqref{eq:max_shell_action}, especially for pure gravity. It turns out that the pure GR action is similar to the free scalar action \eqref{eq:scalar_S_final}, but is even simpler. In particular, it reduces to a sum of codimension-2 integrals, this time without any assumptions on the null boundary's shape. We also identify a universal imaginary term that reproduces the Bekenstein-Hawking entropy. This term is accompanied by a divergence in the action's real part.

\subsection{The contribution from area expansion} \label{sec:gravity:expansion}

We begin with the GR action, complete with the Gibbons-Hawking boundary term \cite{York:1972sj,Gibbons:1976ue}. For non-null boundaries, this reads:
\begin{align}
 S = \frac{1}{16\pi G}\int_\Omega \sqrt{-g}(R + \Lambda + 16\pi G\mathcal{L}_M)\, d^dx 
  + \frac{1}{8\pi G}\int_{\del\Omega} \sqrt{\left|h\right|}K d^{d-1}x \ . \label{eq:GR_action}
\end{align}
In the bulk term, we allowed for a cosmological constant $\Lambda$ and a matter Lagrangian $\mathcal{L}_M$. In the boundary term, 
$h$ is the determinant of the (spacelike or timelike) boundary metric; $K$ is the trace of the extrinsic curvature, defined using the outgoing unit covector (which may correspond to either the outgoing or ingoing normal \emph{vector}, depending on signature). 

We will focus on the boundary term in \eqref{eq:GR_action}. In pure GR, without matter or cosmological constant, this gives the entire action, since the bulk term vanishes on-shell. In any case, we will see that the boundary term has special features - an imaginary part and a divergent real part - that distinguish it from the bulk term.   

Adapting the boundary term to the null setup is a bit non-trivial. Indeed, neither factor in the product $\sqrt{\left|h\right|}K$ has a well-defined finite limit in the null case. The volume density $\sqrt{\left|h\right|}$ vanishes, and there is no unit normal with which to define $K$. However, the product itself does have a null limit. It is given by the area expansion rate $\del_a s^a$, where $s^a$ is the boundary area current. Intuitively, both quantities describe volume or area increase along the boundary's normal. A more rigorous argument is given below.

First, extend the boundary hypersurface into a family of foliating hypersurfaces in a small neighborhood of the point of interest. They should all have the same signature, e.g. in the null case they should all be null. Also, in the non-null case, choose the hypersurfaces to be evenly spaced, so that the normal distance from each to its neighbor is constant. One can then define a coordinate $r$ labeling the hypersurfaces, such that the outgoing covector $N_\mu = \del_\mu r$ has a constant (not necessarily unit) norm. In the null case, that norm goes to zero. As before, we denote the coordinates within each hypersurface by $x^a$. It is now easy to show that in the adapted coordinates $x^\mu = (r,x^a)$, both $\sqrt{\left|h\right|}K$ (in the non-null case) and $\del_a s^a$ (in the null case) are given by $g^{\mu\nu}N_\mu\del_\nu\sqrt{-g}$. We conclude that $\del_a s^a$ is the correct null limit of $\sqrt{\left|h\right|}K$. 

The Gibbons-Hawking integral over the boundary lightsheets therefore reads:
\begin{align}
 S_{exp.} = \frac{1}{8\pi G}\int_{\del\Omega} \del_a s^a d^{d-1}x \ .
\end{align}
This integral is immediately reduced to a linear combination of the lightsheet intersection areas (see section \ref{sec:class:scalar_action} for the sign considerations etc.):
\begin{align}
 \begin{split}
   S_{exp.} &= \frac{1}{4\pi G}\left(\sum_{\mathrm{equators}} - \sum_{\mathrm{tips}}\right)\int \sqrt{\gamma}\, d^{d-2}x \\
     &= \frac{1}{4\pi G}(A_{eq.} - A_{tip,i} - A_{tip,f}) \ , \label{eq:GR_action_exp}
 \end{split}
\end{align}
where $A_{eq.}$, $A_{tip,i}$ and $A_{tip,f}$ are the equator, initial tip and final tip areas, respectively.
 
This remarkable formula, like the free-scalar result \eqref{eq:scalar_S_final}, is a local functional of the boundary configuration fields - in this case, the intrinsic metric. As discussed in section \ref{sec:class:scalar_locality}, in the null case, such an action should not be considered trivial. The additional contributions derived in section \ref{sec:gravity:corner} below will also be linear combinations of the intersection areas. In the jargon of section \ref{sec:class:scalar_holography}, this means that the action of pure GR is ``explicitly holographic''. Its behavior in terms of independent boundary data is much more difficult to study, due to the theory's non-linearity. The reasonable bet for a generic interacting theory is that the action is non-local (section \ref{sec:class:scalar_locality}), while ``local-set holography'' (section \ref{sec:class:scalar_holography}) is absent. Also, the ``explicit holography'' of \eqref{eq:GR_action_exp} is apparently restricted to pure gravity. Generically, we expect that the bulk term in \eqref{eq:GR_action} cannot even be reduced to a codimension-1 integral.  

\subsection{The corner contribution and Bekenstein-Hawking entropy} \label{sec:gravity:corner}

The contribution \eqref{eq:GR_action_exp} arises from area expansion towards the equator. Alternatively, it can be viewed as the bending of the lightsheet normal, as one moves transversely to its generating lightrays. There is no proper sense in which a lightsheet bends \emph{along} its generating rays, since the latter are geodesics. However, this is no longer true when two lightsheets intersect. At the intersection, the normal vector must instantaneously ``bend'' as one passes from one lightsheet to the other. Let us analyze the resulting contribution to the boundary action.

In Euclidean signature, the contribution to the boundary term from hypersurface intersections is well-known \cite{Hartle:1981cf}. Such ``corners'' can be understood as limiting cases of fast-bending smooth curves. This translates into a delta-function in the extrinsic curvature, which has a well-defined integral. The corner contribution to the action is then $\theta A/(8\pi G)$, where $A$ is the codimension-2 area of the ``corner'', and $\theta$ is the bending angle.

\begin{figure}%
\centering%
\includegraphics[scale=0.75]{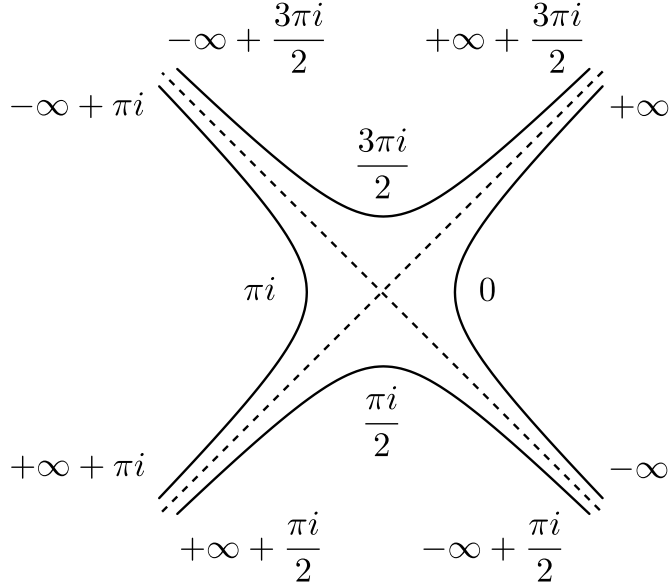} \\
\caption{An assignment of boost angles $\eta$ in the Lorentzian plane, according to eq. \eqref{eq:eta}. This is one of two complex-conjugate choices, distinguished by the sign convention below \eqref{eq:eta}. The horizontal and vertical axes are $x$ and $t$, respectively. The angles are defined up to integer multiples of $2\pi i$.}
\label{fig:angles_plane} 
\end{figure}

In the Lorentzian case, complications arise, since angles (i.e. boost parameters) can now diverge. This is particularly true for null directions. A consistent assignment of angles $\eta$ for the entire 1+1d plane is given in Appendix C of \cite{SorkinThesis}. In each quadrant of the Lorentzian plane, the boost angle runs between $-\infty$ and $+\infty$, diverging at the null lines. When crossing a null line counter-clockwise into a neighboring quadrant, the angle picks up an \emph{imaginary contribution} $-\pi i/2$, without changing the sign of the infinite real part. See figure \ref{fig:angles_plane}. Thus, the total angle for a complete counter-clockwise circuit is $-2\pi i$, in place of the Euclidean $2\pi$. The virtue of this assignment is that $\cosh\eta$ and $\sinh\eta$ are given consistently by:
\begin{align}
 \cosh\eta = \frac{x}{\sqrt{x^2 - t^2}}; \quad \sinh\eta = \frac{t}{\sqrt{x^2 - t^2}} \ , \label{eq:eta}
\end{align}
where we take $\sqrt{s} \equiv i\sqrt{\left|s\right|}$ for negative $s$. In other words, we are analytically continuing \eqref{eq:eta} away from the $x>\left|t\right|$ quadrant. The procedure is unique up to complex conjugation. We will see that the above choice is preferred over its complex conjugate by the requirement that quantum amplitudes $e^{iS}$ do not explode exponentially. Our conventions differ from \cite{SorkinThesis} by an overall factor of $i$. In particular, we treat boost angles within a quadrant as real. 

\begin{figure}%
\centering%
\includegraphics[scale=0.75]{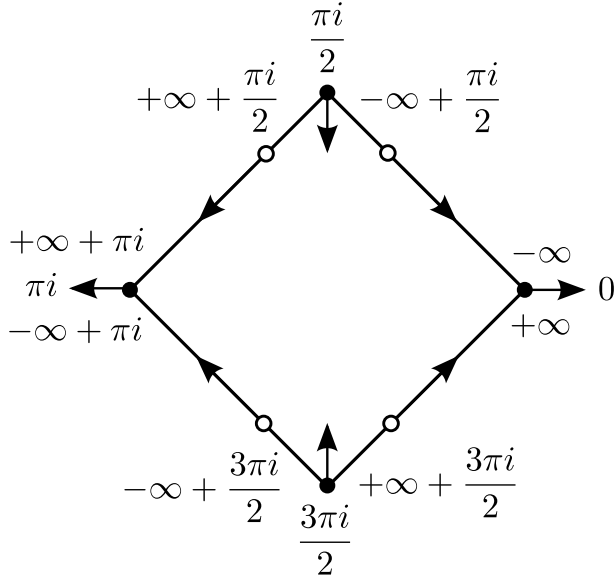} \\
\caption{The boost angles from figure \ref{fig:angles_plane}, as applied to the null boundary normals from figure \ref{fig:boundary}. For clarity, spacelike and timelike normals are included at the equator and tips. They arise if one replaces the corners with smooth curves. Empty circles denote the ``signature flip'' surfaces, where the boost angle crosses into a neighboring quadrant. The path integral favors placing them infinitesimally close to the tips. The ``corner'' contributions to the action can be read off by traveling counter-clockwise and picking up angle differences times $A/(8\pi G)$ factors.}
\label{fig:angles_boundary} 
\end{figure}

Let us apply this recipe to the normal direction on our null boundary and its corners. As with ordinary corners, we can imagine them as fast-bending smooth curves. We then have momentarily a spacelike normal on the equator and timelike normals at the tips. To keep track of the angles, it's best to consider a 1+1d version of the boundary, where the rays actually close into a diamond. See figure \ref{fig:angles_boundary}. The higher-dimensional result is obtained by inserting area factors where necessary. As an intermediate step, one can think of a boundary with $D_1\times\mathcal{M}$ topology (section \ref{sec:why_null:null}). 

Each corner contributes a divergent term to the action, due to the infinite boost between one null direction and the other. The sum of these divergent terms reads:
\begin{align}
 S_{boost} = -\frac{\theta}{8\pi G}(A_{eq.} - A_{tip,i} - A_{tip,f}); \quad \theta \rightarrow +\infty \ . \label{eq:GR_action_boost}
\end{align}
This is the same as $S_{exp.}$ from \eqref{eq:GR_action_exp}, but with a factor of $-\infty$. The signs in \eqref{eq:GR_action_boost} can be read off from the angle differences in figure \ref{fig:angles_boundary}, as one walks along it clockwise. This can be verified by evaluating $K$ on smooth ``zoomed-in'' versions of the equator and tips, i.e. on convex timelike and spacelike curves, respectively. We note that when $A_{eq.} = A_{tip,i} + A_{tip,f}$, as is always the case in 1+1d, $S_{boost}$ may have a finite value.

The normal angle should also get imaginary contributions of $-\pi i/2$ at signature flips, i.e. when it crosses into a different quadrant of the Lorentzian plane. All we can say from figure \ref{fig:angles_boundary} is that this happens \emph{somewhere} along the null extent of each lightray. Overall, there should be two surfaces, one on the initial half-boundary and one on the final half, where the signature flip takes place. Let us denote their areas by $A_{flip,i}$ and $A_{flip,f}$, respectively. Applying the same counter-clockwise recipe for the signs, we find the following imaginary contribution to the action:
\begin{align}
 S_{flip} = \frac{1}{8\pi G}\cdot \frac{\pi i}{2}(A_{flip,i} + A_{flip,f}) = \frac{i}{16G}(A_{flip,i} + A_{flip,f}) \ . \label{eq:GR_action_flip}
\end{align}

The full boundary term in \eqref{eq:GR_action} is given by $S_{exp.} + S_{boost} + S_{flip}$. The bulk term is generically real and finite on-shell. Therefore, $S_{boost}$ is the divergent part of the total action, while $S_{flip}$ provides the imaginary part. The complete GR action thus satisfies:
\begin{align}
 \Im S = \frac{1}{16G}(A_{flip,i} + A_{flip,f}) \ . \label{eq:Im_S_raw}
\end{align}           

It remains to fix $A_{flip,i}$ and $A_{flip,f}$. To do this, we can imagine a path integral over the different possibilities. The imaginary action $\Im S$ suppresses the amplitudes $e^{iS}$ exponentially, as $e^{-\Im S}$. Like in Euclidean path integrals, this selects the smallest possible imaginary action. Thus, we should give $A_{flip,i}$ and $A_{flip,f}$ the smallest possible values. Now, it is known \cite{Bousso:1999xy} that on-shell, the boundary area decreases monotonously from the equator towards the tips (and will continue decreasing until a caustic is reached). Thus, the ``flip'' surfaces must be chosen next to the tips. This fixes $A_{flip,i} = 2A_{tip,i}$ and $A_{flip,f} = 2A_{tip,f}$. The factor of 2 arises from the two rays intersecting at each tip point. Alternatively, the boundary area infinitesimally close to the tip is twice the tip area. Substituting into \eqref{eq:Im_S_raw}, we arrive at the following result in terms of the \emph{average} tip area:
\begin{align}
  \Im S = \frac{1}{4G}\cdot\frac{A_{tip,i} + A_{tip,f}}{2} \ . \label{eq:Im_S}
\end{align}
This is remarkably similar to the Bekenstein-Hawking entropy formula. Indeed, an imaginary (or Euclidean) action is usually identified with the logarithm of a partition function. This is the same as the entropy, up to terms arising from conserved charges, such as $E/T$ for energy and $J\Omega/T$ for angular momentum. For a Euclidean black hole, the latter can be viewed as contributions from the boundary at infinity, while the entropy is a contribution from the horizon \cite{Carlip:1993sa}. This might explain why the imaginary action \eqref{eq:Im_S} gives the entropy formula directly: for our boundary structure, there is no ``term at infinity'' to consider.

It should be noted that contributions of the form \eqref{eq:GR_action_boost}-\eqref{eq:GR_action_flip} are not unique to null boundaries. Any closed hypersurface in spacetime contains infinite boosts and signature flips. The consequences for more general boundaries are addressed in a separate work \cite{Neiman:2013ap}. It turns out that the action's imaginary part remains related to the black hole entropy formula, including in higher-order Lovelock gravity. This coincidence was explained in \cite{Neiman:2013lxa} by relating the Lorentzian calculation to a Euclidean one.

\section{Discussion} \label{sec:discuss}

In this paper, we argued that restricting to null boundaries may be a useful step in the grand project of finite-region quantum gravity. We set out to explore this idea in the context of classical field theory. The expected breakdown of conventional observables in finite-region QG is then mirrored by a breakdown of the standard tools of classical mechanics. Somewhat fancifully, we identified certain artifacts in the null version of classical field theory with certain features that are expected of quantum gravity - non-locality, holography and the absence of global charges. Finally, while exploring the GR action for null boundaries, we found an imaginary term that resembles the Bekenstein-Hawking entropy.          

In our interpretation of the on-shell actions, we claimed that terms such as \eqref{eq:scalar_S_final} and \eqref{eq:Im_S}, which are local in configuration data, must be taken seriously in the null case, and should in fact be treated as ``non-local''. A possible objection is that such terms can be added to the action \emph{before} taking the null limit, rendering them ambiguous. However, it seems hard to create such ambiguity in practice. To be concrete, let us demand that any new boundary term must be the null limit of a well-defined, analytical expression in the non-null case. For classical field theory, this looks like a sensible requirement. First, note that \eqref{eq:scalar_S_final} and \eqref{eq:Im_S} themselves arise from terms that involve normal derivatives. Outside the null limit, those are \emph{not} local in configuration data, and are therefore not arbitrary. If we try to add boundary terms \emph{without} normal derivatives, there appear to be two possible outcomes. They either vanish in the null limit (due to the vanishing volume density), or become ill-defined (due to factors of the inverse boundary metric). One may also consider introducing codimension-2 terms such as \eqref{eq:scalar_S_final} or \eqref{eq:Im_S} directly into the non-null action, by assigning them to corners, or to loci of signature flips. However, such singling out of special surfaces in the general action formula can be ruled out by analyticity.

The most significant result of this work is arguably the GR ``entropy'' formula \eqref{eq:Im_S}. Its derivation bears some resemblance to the analysis of Euclidean black holes \cite{Gibbons:1976ue}. There as well as here, an imaginary action is derived through analytical continuation. However, in our derivation, there is no Wick rotation of the entire metric, and thus no stationarity requirement. Instead, one can view the imaginary boosts during signature flips as resulting from ``localized Wick rotations''. Another distinguishing feature of our calculation is that it makes no reference to asymptotic infinity. The entropy is obtained directly, without involving thermodynamic potentials (associated with stationarity) or conserved charges (associated with boundary terms at infinity).

A curious feature of the ``entropy'' result \eqref{eq:Im_S} is that the relevant areas are those of the \emph{smallest} boundary sections. This contrasts with the notion that the entropy bound is concerned with the largest available area, in particular in the null context \cite{Bousso:1999xy}. The picture suggested by \eqref{eq:Im_S} seems more compelling: the smallest area defines a bottleneck on the available information. For instance, a causal diamond with vanishing tip areas $A_{tip,i} = A_{tip,f} = 0$ is associated with a pointlike observer \cite{Bousso:2000nf}. Now, a truly pointlike observer cannot carry any information at all. In particular, he cannot design an experiment in his causal diamond, or collect the results of measurements. From this perspective, it stands to reason that the entropy bound for a causal diamond should be zero.

An important sanity check for our ``entropy'' calculation is to apply it to a stationary black hole solution. This turns out to be possible, and produces the correct result \cite{Neiman:2013lxa}.  

We conclude with a comment on the nature of the quantum theory which was advocated in section \ref{sec:why_null}. One's first impulse would be to assign a Hilbert space to the initial and final half-boundaries, and to think of a path integral as providing an ``S-matrix'' of transition amplitudes. However, this seems problematic. The issue is with the operational meaning of Hilbert products between states on the same hypersurface. In ordinary QFT, they can be viewed as a limiting case of transition amplitudes between states on different hypersurfaces, obtained as the hypersurfaces approach one another. In the null setup, this procedure fails, due to the rigidity of lightsheets: one cannot have two infinitesimally close lightsheets with support on the same equator. There are only transition amplitudes between the initial and final lightsheets, with no intermediates. These amplitudes define a map linking the initial and final states, but not a Hilbert product for either. In \cite{Witten:2001kn}, a similar situation was described, with the past and future infinities of de-Sitter in place of our initial and final lightsheets. There, a Hilbert product was constructed by using also the CPT map between initial and final states. This does not seem to work in our setup, since an irregular equator would not be invariant under CPT. This failure in constructing a Hilbert space is consistent with the expected failure of standard quantum mechanics in finite-region QG. The nature of the theory's correct output is an open question. A variant of these remarks applies also to LQG and spinfoams. There, one cannot understand the Hilbert product as a limit of transition amplitudes because the boundary hypersurfaces are \emph{too free} to move around: the initial and final half-boundaries can approach each other in principle, but it isn't clear how to \emph{make} them approach each other using GR-like boundary data. 

\section*{Acknowledgements}		

This work is supported in part by the NSF grant PHY-1205388 and the Eberly Research Funds of Penn State.
The early stages were carried out in Tel Aviv University, supported in part by the Israeli Science Foundation center of excellence, the US-Israel Binational Science Foundation (BSF), the German-Israeli Foundation (GIF) and the Buchmann Scholarship Fund.

\end{document}